\documentclass{ws-procs975x65}

\newcommand{\da}{\partial}
\newcommand{\E}{\mathcal{E}}

\begin{document}

\title{Spherically Symmetric Gravitational Collapse of Perfect Fluids}

\author{P. LASKY and A. LUN}

\address{School of Mathematical Sciences, Monash University,\\
Melbourne, Victoria 3800, Australia\\
E-mail: paul.lasky@sci.monash.edu.au\\
E-mail: anthony.lun@sci.monash.edu.au}

\begin{abstract}
Formulating a perfect fluid filled spherically symmetric metric utilizing the 3+1 formalism for general relativity, we show that the metric coefficients are completely determined by the mass-energy distribution, and its time rate of change on an initial spacelike hypersurface.  Rather than specifying Schwarzschild coordinates for the exterior of the collapsing region, we let the interior dictate the form of the solution in the exterior, and thus both regions are found to be written in one coordinate patch.  This not only alleviates the need for complicated matching schemes at the interface, but also finds a new coordinate system for the Schwarzschild spacetime expressed in generalized Painleve-Gullstrand coordinates. 
\end{abstract}


\bodymatter

\section*{}
The traditional approach to the analysis of gravitational collapse follows that devised by Oppenheimer and Snyder\cite{oppenheimer39}, whereby the Einstein field equations are solved for the interior, matter-filled region without consideration of the exterior.  Whence a solution to the interior is found, the Israel-Darmois matching conditions are utilized to ``glue'' the interior spacetime to an appropriate exterior spacetime, commonly Schwarzschild.  Throughout the process the two regions are considered as separate entities, mainly as they are described by different coordinate systems.

We introduce a different approach, whereby the spacetime is established as an initial/boundary value problem, with the interface between the two regions of the spacetime being a free-boundary.  This enables us to describe both regions of the spacetime under a single coordinate patch by simply letting the energy-momentum variables go to zero at some finite coordinate radius.  In this talk we use our formalism to describe the gravitational collapse of a spherically symmetric perfect fluid.  

As we are setting up an initial value problem, an ideal starting point is the ADM system of equations.  This will enable us to establish an initial spacelike hypersurface, with perfect fluid for $r\le r_{\partial}$ and vacuum for $r>r_{\partial}$, where $r_{\partial}$ is some radius on the initial slice.  The system can then be evolved forward in time to describe the entire collapse process.  Furthermore, in order to describe both regions of the spacetime utilizing a single coordinate system, one requires that the observer have a finite radial velocity such that this observer will pass through the interface between the two regions.  In ADM language, this implies a non-zero, radial component of the shift vector, $\beta(t,r)$.  We therefore begin with an arbitrary, spherically symmetric line element expressed as
\begin{align}
d\mathcal{S}^{2}=-\alpha^{2}dt^{2}+\left(1+\E\right)^{-1}\left(\beta dt+dr\right)^{2}+r^{2}d\Omega^{2},        
\end{align}
where $\alpha(t,r)$ is the lapse function, $\E(t,r)>-1$ is an arbitrary function which reduces to the energy function of the Lemaitre-Tolman (LT) metric\cite{lemaitre33,tolman34}, and $d\Omega^{2}$ is the metric of the two sphere.  

By putting this line element through the ADM equations (for details see  \cite{lasky06,lasky06a}), one can derive the reduced field equations which are a coupled system of first order differential equations.  We define a ``mass'' function\footnote{We note in the dust limit, $M$ becomes the familiar mass of the LT solution, and in the vacuum limit becomes the Schwarzschild mass.}
\begin{align}
M:=4\pi\int_{s=0}^{r}\rho(t,s)s^{2}ds,\label{mass}
\end{align}  
where $\rho(t,r)$ is the mass-energy density.  The lapse function is related to the density and pressure through Euler's equation
\begin{align}
\da_{r}P=-\left(\rho+P\right)\da_{r}\left(\ln\alpha\right).\label{Euler}
\end{align}  
The solution of this equation requires the specification of an equation of state (EoS) which relates the density to the pressure, $P$.  Thus, given an EoS, the system reduces to the line element along with two equations,
\begin{align}
d\mathcal{S}&^{2}=\left(1+\E\right)^{-1}\left[-\alpha^{2}\left(1-\frac{2M}{r}\right)dt^{2}+2\alpha\sqrt{\frac{2M}{r}+\E}\,\,dtdr+dr^{2}+r^{2}d\Omega^{2}\right],\label{lineelement}\\
&\mathcal{L}_{n}\E=\frac{2\left(1+\E\right)}{\rho+P}\sqrt{\frac{2M}{r}+\E}\,\,\da_{r}P\qquad\textrm{and}\qquad\mathcal{L}_{n}M=4\pi Pr^{2}\sqrt{\frac{2M}{r}+\E}.\label{ev}
\end{align}
Here, $\mathcal{L}_{n}$ denotes the Lie derivative with respect to the unit normal vector which, when acting on a scalar, takes the form $\mathcal{L}_{n}\psi=\alpha^{-1}\left(\da_{t}\psi-\beta\da_{r}\psi\right)$.  Once an EoS is specified, equation (\ref{Euler}) is solved and hence the lapse function can be written in terms of the density, and thus the mass.  Therefore, equations (\ref{ev}) are two equations for two unknown functions $M$ and $\E$.  

As per our aim, these equations describe both the interior perfect fluid region and the exterior vacuum region in the one coordinate patch.  To see this we simply let the pressure and density vanish external to some finite radius.  Equation (\ref{mass}) then implies that the mass function is constant, which further implies the right hand equation in (\ref{ev}) is trivially satisfied.  Equation (\ref{Euler}) implies the lapse function is simply a function of the temporal coordinate, and utilizing coordinate freedom the lapse can be set to unity without loss of generality.  The resulting system is what we call the generalized Painleve-Gullstrand (GPG) line element as the special case ($\E=0$) is the Painleve-Gullstrand line element.  

The GPG class of solutions comprise a family of coordinate systems for the Schwarzschild spacetime.  The coordinate transformation between this class of solutions and Schwarzschild coordinates, $\left(\hat{t},r,\theta,\phi\right)$, is given by the solution to the coupled differential equations
\begin{align}
\left(\da_{\hat{t}}t\right)^{2}=1+\E\qquad\textrm{and}\qquad\left(1-\frac{2M}{r}\right)\da_{r}t=\sqrt{\frac{2M}{r}+\E},
\end{align}
where $t=t(\hat{t},r)$.  One can show a solution exists to these equations, and thus the coordinate transformation is always valid by utilizing the integrability conditions, which are satisfied providing the left hand equation in (\ref{ev}) is satisfied (for details see \cite{lasky06}). 

Reverting back to the full system of equations in the interior with matter (\ref{mass}-\ref{ev}), we can transfer these into diagonal coordinates, $\left(\tau,R,\theta,\phi\right)$, such that the generalization from the LT metric for dust becomes obvious.  By letting $r=r(T,R)$ such that
\begin{align}
\left(\da_{\tau}r\right)^{2}=\alpha^{2}\left(\frac{2M}{r}+\E\right),
\end{align}
the line element and equations (\ref{ev}) become
\begin{align}
&d\mathcal{S}^{2}=-\alpha^{2}d\tau^{2}+\frac{\left(\da_{R}r\right)^{2}}{1+\E}dR^{2}+rd\Omega^{2},\\
\left(\da_{R}r\right)\left(\da_{\tau}\E\right)=\frac{1+\E}{\rho+P}&\alpha\sqrt{\frac{2M}{r}+\E}\,\,\da_{R}P\,\,\,\,\,\textrm{and}\,\,\,\,\, \da_{\tau}M=4\pi P r^{2}\alpha\sqrt{\frac{2M}{r}+\E},
\end{align}
and equations (\ref{mass}) and (\ref{Euler}) are suitably dealt with.  The reduction to the LT dust models is clear, again using the coordinate freedom that the lapse function becomes a function of time, and can thus be set to unity without loss of generality. 

A number of extensions of this work are currently under investigation:
\begin{itemize}
  \item Searching for exact solutions of equations (\ref{mass})-(\ref{ev}).
  \item The analysis and determination of shell-crossing singularities which exhibit themselves as fluid shock waves in this coordinate system.  
  \item The relaxation of the perfect fluid condition to allow for more realistic matter sources, enabling the study of diffusion processes and anisotropic stresses.
  \item A relaxation of the symmetries of the geometry to allow for quasi-spherical symmetry or axial-symmetry.   
\end{itemize}

\bibliographystyle{ws-procs975x65}
\bibliography{Bib}

\end{document}